# Geometrically Enhanced Thermoelectric Effects in Graphene Nanoconstrictions

*Achim Harzheim,[†, ||] Jean Spiece,[‡, ||] Charalambos Evangeli,[†,‡] Edward McCann,[‡] Vladimir Falko,[§] Yuewen Sheng,[†] Jamie H. Warner,[†] G. Andrew D. Briggs,[†] Jan A. Mol,[†] Pascal Gehring,[†,¶,*] and Oleg V. Kolosov,[‡, *]*

[†] Department of Materials
University of Oxford
Parks Road, OX1 3PH Oxford, UK

[‡] Department of Physics
Lancaster University
Bailrigg, LA1 4YB, Lancaster, UK

[§] School of Physics & Astronomy
University of Manchester
Oxford Road, M13 9PL, Manchester, UK

[¶] Kavli Institute of Nanoscience
Delft University of Technology
Lorentzweg 1, 2628 CJ, Delft, Netherlands

*Email: p.gehring@udelft.nl; o.kolosov@lancaster.ac.uk

ABSTRACT. The influence of nanostructuring and quantum confinement on the thermoelectric properties of materials has been extensively studied. While this has made possible multiple breakthroughs in the achievable figure of merit, classical confinement and its effect on the local Seebeck coefficient has mostly been neglected, as has the Peltier effect in general due to the complexity of measuring small temperature gradients locally.

Here we report that reducing the width of a graphene channel to 100 nanometers changes the Seebeck coefficient by orders of magnitude. Using a scanning thermal microscope allows us to probe the local temperature of electrically contacted graphene two-terminal devices or to locally heat the sample. We show that constrictions in mono- and bilayer graphene facilitate a spatially correlated gradient in the Seebeck and Peltier coefficient, as evidenced by the pronounced thermovoltage $V_{th}$ and heating/cooling response $\Delta T_{Peltier}$ respectively. This geometry dependent effect, which has not been reported previously in 2D materials, has important implications for measurements of patterned nanostructures in graphene and points to novel solutions for effective thermal management in electronic graphene devices or concepts for single material thermocouples.

TEXT. Solid-state thermoelectric devices have long been attractive to researchers and engineers alike, due to their capability of reliably converting waste heat to electricity and the possible thermal management applications.[1–4] In addition, an in-depth understanding of thermoelectric phenomena is important to correctly interpret photocurrent and electrical transport measurements where these phenomena can play a major role.[4,5] There are two complementary thermoelectric effects, the Seebeck effect and its Onsager reciprocal, the Peltier effect. For the first, a temperature difference $\Delta T$ will induce the buildup of a thermovoltage $\Delta V = -S\Delta T$ across a material with a Seebeck coefficient $S$. Vice versa, for the second, an electrical current $I$ induces a heat flow $\dot{Q} = \Pi I$, where $\Pi = TS$ is the Peltier coefficient.[6]

A resurge in interest in this topic was initiated by Hicks and Dresselhaus' theoretical findings that reducing the dimensionality of thermoelectric materials could significantly increase their efficiency.[7,8] This is measured by the dimensionless figure of merit $ZT = \frac{S^2\sigma}{\kappa}T$ – a function of the electrical ($\sigma$) and thermal ($\kappa$) conductivity – and the principle has since been demonstrated by various groups.[9,10] Amongst the techniques that have been employed are building nanocomposites from nanocrystal blocks,[11] nanostructuring quantum dot superlattices,[9] the exploitation of negative correlations between electrical and thermal conductivity,[12] and band engineering.[13,14] Moreover, classical rather than quantum confinement has been reported to cause an increase in the Seebeck coefficient in gold and Antimony Telluride nanowires.[15,16]

Here we present high resolution Scanning Thermal Microscopy measurements of 100 nm wide graphene bow-tie nanoconstrictions that show a pronounced spatial dependence of the Seebeck and the Peltier effect. This change in the local Seebeck coefficient is attributed to a shortened effective Electron Mean Free Path (EMFP) due to edge scattering and opens up the possibility to readily produce two dimensional one-material thermocouples as well as accessible local temperature management and improved heat dissipation.

We perform our measurements with a Scanning Thermal Microscope (SThM) – effectively an atomic force microscop (AFM) with a microfabricated resistor incorporated close to the tip[17] - using two different protocols to map the Seebeck and Peltier effect as well as the Joule heating. In the Peltier measurement, we use a recently developed non-equilibrium scanning probe thermometry method:[18] an AC bias $V_{\text{bias}}$ applied to the device through the global contacts causes an AC current $I_{\text{bias}}$ which results in Joule heating and Peltier heating/cooling. By measuring the temperature response of our SThM tip as it is scanned over the AC biased sample and modulating it at the first (Peltier) and second (Joule) harmonic it is possible to decouple the two effects and extract the respective heating/cooling values (see Figure 1a for the measurement schematics). In contrast, for the thermovoltage or Seebeck measurement, the SThM tip is heated by applying a high AC voltage to it and the global voltage drop over the device is recorded at the second harmonic as the hot tip is scanned over the sample. Both single layer and multilayer graphene are measured, but no thickness dependence in the size and distribution of the signal is found.

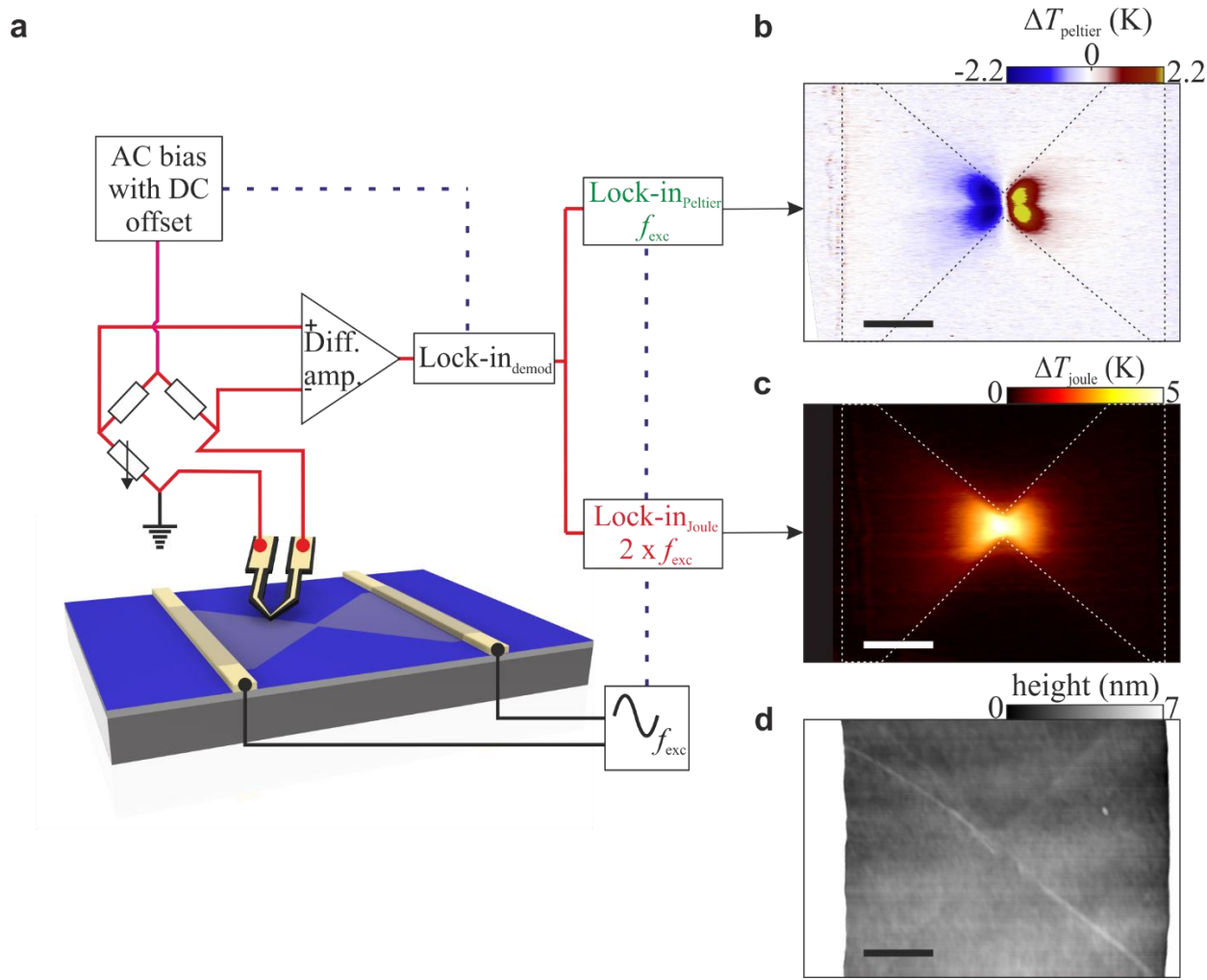

**Figure 1.** Nanoscale mapping of the Peltier effect in graphene nanoconstrictions. (a) An AC voltage bias $V_{bias}$ at $f_{exc}$ induces an AC current $I_{bias}$ through the constriction (black lines). In addition, a low AC bias with a DC offset is applied to the SThM tip through a Wheatstone bridge (magenta line). During scanning, the resulting signal in the tip (red lines) is demodulated at the respective frequency. This thermal signal is then demodulated at the first ($f_{exc}$) and second ($2f_{exc}$) harmonic, providing the Peltier heating/cooling (green) in b and Joule heating (red) in c, respectively. The blue dashed lines symbolize the reference signal lines. (b) Peltier effect map showing the main heating/cooling effects around the constriction (c) Joule heating map, showing the hot spot in the middle of the constriction. (d) Simultaneously recorded height map used to outline the position of the constriction in the Peltier heating/cooling and Joule heating images. In b and c, the dotted-dashed lines indicate the contact position and the dashed line the outline of the graphene constriction. All scale bars are 1 µm.

Peltier and Joule heating maps of the bow tie device, are shown in Figure 1b and c, respectively, where both show a high spatial dependence, with a strong signal around the constriction. The Joule heating exhibits a temperature increase while the Peltier signal shows cooling/heating on the respective side of the constriction and a node in the middle. The Peltier signal shown here corresponds to the measured amplitude multiplied by the sine of the phase signal. It is the temperature at a certain phase at the maximum applied modulation voltage. It is worth to mention that in time average no discernible Peltier heating or cooling is taking place at the constriction for an AC bias.[18]

Figure 1d shows the simultaneously measured height signal, which was used to determine the exact position of the device indicated in b and c.

The Joule heating showing a maximum in the constriction is expected due to the increased local current density,[19] however, given the continuous composition of the material in the constriction area, all thermoelectric effects in the device would be expected only in the vicinity of the Au electrodes.[20] As can be seen in Figure 1b, the Peltier signal $\Delta T_\text{Peltier}$ becomes strongest around the constriction itself and outlines the shape of the graphene bow-tie where the signal at the edges is broadened out due to heat spreading into the surrounding $SiO_2$ substrate. The SThM measurement of the device without current excitation, shows that the heat dissipation from the heated tip in the areas with and without graphene differs by less than 5% (see SI section 7). This is in agreement with findings by Tovee *et al.*[21] on SThM scanning of solid state materials. Thus, it is reasonable to assume that the heat mostly spreads in the $SiO_2$/Si substrate. The Peltier effect results in heating and cooling of up to $\Delta T_\text{Peltier} \approx \pm 2K$ on either side of the constriction for an applied current of $I_{bias} \approx 90\mu A$. A markedly similar behavior was found for $V_\text{th}$ in the thermovoltage measurements on the same device (see Figure S5 in the Supporting Information) under open-circuit condition, confirming that the signal likely stems from a changed local Seebeck coefficient. In addition, we observe comparatively weak "conventional" Peltier heating/cooling in the vicinity of the Au contacts (see Figure 1b) which is explained by the formation of a Peltier junction between gold and graphene at the contacts as reported previously.[4,20]

Such a geometrical modification of the local Seebeck coefficient has been seen in metallic thin-film stripes and Au nanowires and was explained by structural defects and the metal grain structure, which in turn reduce the EMFP.[15,22] The EMFP of graphene at room temperature, is typically on the order of 100s of nanometers and thus higher than in gold.[23] However, it gets substantially reduced by defect potentials such as ones stemming from rough edges,[24] which in our case have been created by the device patterning and amount up to an 80% reduction.[25] This edge scattering becomes more dominant as the width of the graphene stripe $\Delta y(x)$ reduces, giving a position dependent mean free path, which can be written as

$$l(x) = l_0 \left[1 + c_n \left(\frac{l_0}{\Delta y(x)}\right)^n\right]^{-1}, \qquad (1)$$

where $l_0$ is the bulk mean free path and $c_n$ and $n$ are numerical coefficients specifying the transport mode and the influence of scattering on the mean free path (see section 11 in the Supporting Information). To extract the bulk mean free path we perform gate conductance measurements on 43 μm long and 3 μm wide graphene ribbons that give us $l_0 \approx 226 \pm 20$ nm (see section 3 Supporting Information).

Using the Mott formula $S = \frac{\pi^2 k_B^2 T}{3e} \frac{1}{R(\epsilon)} \frac{dR(\epsilon)}{d\epsilon}|_{\epsilon=\epsilon_F}$ we obtain an expression for the thermopower as a function of constriction width (see section 11 in the Supporting Information for more information):

$$S = -\frac{\pi^2 k_B^2 T}{3\epsilon_F e}\left[1 + n\, U \frac{l(x)}{l_0} - (n-1)U\right], \qquad (2)$$

where $U = \frac{d \ln l_0}{d \ln \epsilon}|_{(\epsilon=\epsilon_F)}$ is the exponent of any power law dependence of the EMFP on energy. We expect this value to be between the short range disorder or electron-phonon interaction value $U = -1$ and the long range Coulomb interaction $U = +1$.[26,27] Equation (2) predicts that the local Seebeck coefficient decreases when the width of the channel is reduced, which leads to regions with different effective Seebeck coefficients in the bow-tie shaped devices.

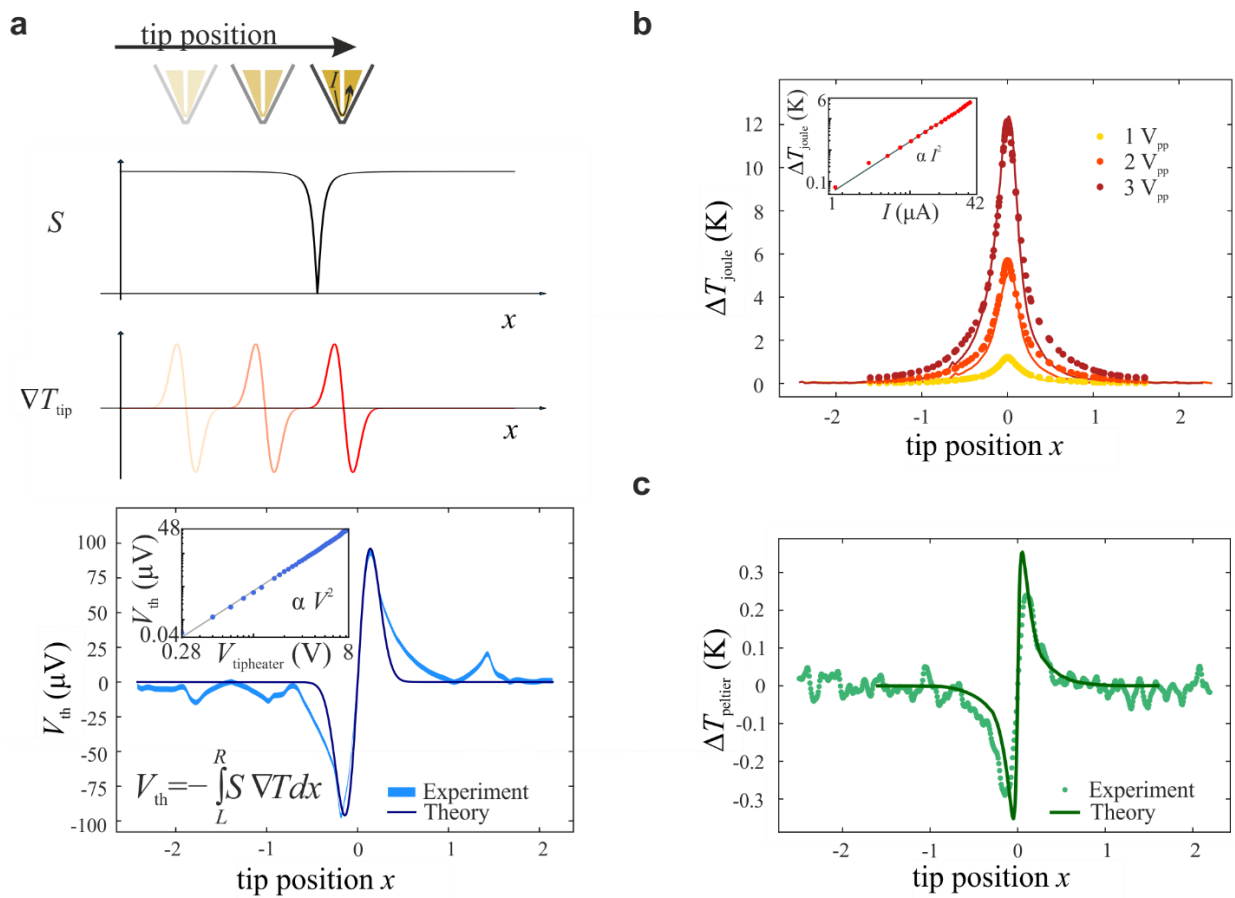

**Figure 2.** Modelling and fitting of Joule heating and thermoelectric effects in a bow-tie device. (a) From the top: schematic of the tip movement, 1D section cuts through the middle of the constriction of the calculated Seebeck coefficient, the tip-defined moving thermal gradient and the resulting thermovoltage measured and calculated respectively. The inset shows the quadratic tip voltage dependence of the thermovoltage signal in a log-log plot. (b) Joule heating at different applied voltage biases experimentally recorded (dots) and fitted to a COMSOL model (lines). The smallest Joule heating signal (1 $V_{pp}$, yellow) is used to extract the electrical and thermal conductivities for the entire model ($\kappa = 120\,\text{Wm}^{-1}\text{K}^{-1}$, $\sigma = 5 \cdot 10^5\,\text{Sm}^{-1}$). (c) Peltier heating/cooling at 1 $V_{pp}$, experimental and simulated from the COMSOL model using the calculated Seebeck coefficient from (a). The zero of the tip position is centered at the middle of the constriction for all figures.

Using Equation (1) and (2), we can model $V_{\text{th}}(x)$ and compare it to the measured thermovoltage 1-D line section signals. As shown in Figure 2a, $V_{\text{th}}(x) = -\int_L^R S\nabla T\, dx$ is calculated by taking the integral of $S\nabla T$ over the whole length of the device at each point. In the measurement and in our calculations, the Seebeck coefficient is only dependent on the width of the constriction and its distribution does not change as we move the tip, while the temperature gradient $\Delta T_{\text{tip}}$ induced by the heater voltage $V_{\text{heater}}$ is always centered at the tip position $x$ and thereby moves as we scan over the sample. The heater temperature $\Delta T_{\text{tip}}$ is obtained from calibrating the tip and measuring the thermal resistance between the heater and the sample (see Supporting Information 7). It is worth noting here that there is an inherent uncertainty of 15-20% of the heater temperature that can lead to an over or underestimation of the measured effect. However this does not change the conclusion and main results of our work. Fitting the calculated values to the line cut of the thermovoltage measured with and estimated $\Delta T_{\text{tip}} \approx 18 \pm 2$ K gives the dimensionless parameters $c_n \approx 0.56$, $n \approx 2.6$ and $U \approx 0.88$. Using these fitting results we calculate a bulk Seebeck coefficient of $S \approx 118$ µV K$^{-1}$, which is similar to values previously found for graphene at room temperature.[28] This value reduces to $S_{\text{min}} \approx 0.34\ \mu V K^{-1}$ in the middle of the junction due to the reduction of the mean free path within the constriction. This decrease by orders of magnitude can be explained by Equation (2): it involves a difference of terms, which results in a big variation of $S$ for relatively small changes in the EMFP.

To further test the influence of geometrical confinement on the thermoelectric properties of graphene devices we have tested an "island" structure, where wide and narrow parts of graphene alternate and which is showing a pronounced signal at these junctions (see Figure S13 in the Supporting Information). It is worth to mention that applying a back gate voltage enables us to change the doping from $p^{++}$-doping (-30V) to $p$-doping (30V) which results in a modification of the signal strength in the constriction by approximately 20% due to the changed carrier density (see Figure S11 in the Supporting Information).[28]

The spatially dependent Seebeck coefficient extracted from the thermovoltage fit can be used to develop a COMSOL model that can reproduce our experimental Joule heating and Peltier signal (see Figure 2b and c). In this model the effective thermal conductivity $\kappa = 120$ W(mK)$^{-1}$ and the electrical conductivity $\sigma = 5 \cdot 10^5$ Sm$^{-1}$ are the only fit parameters. The spatial heat distribution is mainly determined by the SiO$_2$ layer and only slightly modified by the thermal conductivity of the single graphene layer.

We have in addition studied the current dependence of all measured signals by placing the tip on one side of the constriction as the current through the device, $I_{bias}$, (in the Peltier and Joule heating measurements) or through the tip, $I_{heater}$, (in the thermovoltage measurements) is increased. In both the Joule heating and the thermovoltage measurements, a square current dependence on the current is observed (see Figure 2a and b inset), in agreement with the Joule-Lenz law ($P \propto I^2 R$).

However, in the Peltier measurement of the bow-tie device, we find that an initially linear dependence changes to a cubic one as we increase the current $I_{bias}$. As can be seen in Figure 3a, the data can be fitted with a combination of a cubic and linear term, where the crossover point is located at approximately $I_{bias} = 33$ µA. This is a deviation from the simple linear dependence predicted by $\dot{Q} = \Pi * I_{bias} = ST * I_{bias}$. We find this behavior in all geometries measured, with the crossover happening at different current levels (Figure S7 in the Supporting Information).

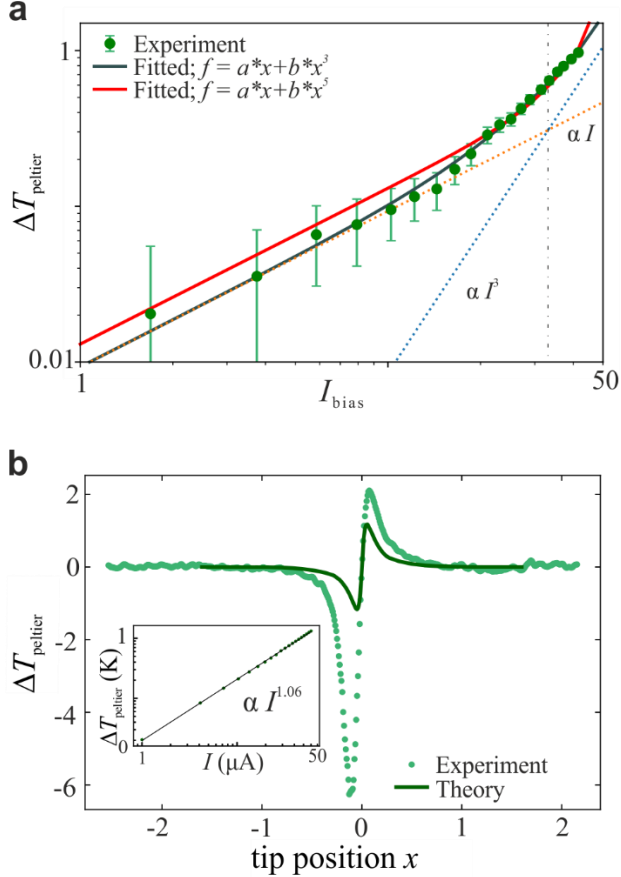

**Figure 3.** Deviation of the experimental data from the linear Peltier model. (a) Fit of the current dependency of the Peltier heating in the constriction for a linear and cubic (grey line) and fifth order term (red line). For the cubic dependency, which seems to fit the data better, the Peltier heating switches over from a linear to a cubic current dependency where the switchover point is marked by the black dotted-dashed line. The orange dotted line is linear with respect to the current and the blue dotted line is cubic and serve as a guide to the eye. (b) Comparison of the Peltier heating/cooling to the COMSOL model at $3V_{pp}$. A big discrepancy between the COMSOL model and the experimental data is visible both in shape and in amplitude. The asymmetry in the experimental data is sample specific and might be linked to the nanoscale structure of the nanoconstriction (see Table T1 in the SI). The inset shows the current dependency of the simulated Peltier heating, which is linear, save for a small correction ($\propto 1.02$) due to the Joule heating.

We attribute the unusual current dependence observed in our experiments to an "electron wind" effect: if the drift velocity $v_{\text{drift}}$ becomes comparable to the Fermi velocity $v_F$ heat is shifted with respect to the position of the constriction, effectively cooling one side and heating the other side. For this effect, we expect the Peltier heating/cooling to take the form of a sum of the common linear Peltier effect and a cubic term. The latter originates from an increasing drift velocity (linearly increasing with current) and the temperature of the hot carriers created by Joule heating (quadratic current dependency) which add up to an additional cubic term (see Supporting Information for a full derivation). Indeed, we find that a fit of this model to the measured data provides a good agreement, compared to other higher order terms (see Figure 3 and SI). The drift velocity in our devices is given by $v_{\text{drift}} = \frac{I}{neW} \approx 0.25 \cdot 10^6 \text{ms}^{-1}$ where $I \approx 40$ μA is the current through the device, $n$ the carrier density, $e$ the elementary charge and $W = 100$ nm the width of the constriction. This velocity is approaching the Fermi velocity in graphene, $v_{\text{F}} \approx 10^6$ ms$^{-1}$. The carrier density is approximated by the low current value of $n = 10^{16}$ m$^{-2}$. A similar electron wind effect has been observed for varying gate voltages in graphene devices.[29]

An alternative origin of the non-linearity of the Peltier effect is the temperature dependence of the Seebeck coefficient. The latter increases because of local Joule heating, which would give rise to a fifth order current dependence since $\dot{Q} \propto STI \propto T^2 I \propto I^5$. However, since the measurements are performed at room temperature (300K) and only a few Kelvin temperature increase due to Joule heating are measured the impact on the Peltier heating/cooling is negligible (see section 1 in the Supporting Information). Furthermore, the temperature increase is also taken into account in the finite element analysis (FEA) of the Peltier heating/cooling, which solves for the full thermoelectric equation ($\rho C_p \boldsymbol{u} \nabla T = \nabla(k \nabla T - P \boldsymbol{J}) + Q$, (see Supporting Information). The results of this FEA suggest a small deviation of only 2% from the linear exponent (see inset Figure 3b), which is about two orders lower than the observed change to a higher order exponent in the experiment.

Nevertheless, it is important to stress, that while a heat drift in the constriction due to high drift currents can explain the observed deviation from a linear current dependency as expected for the Peltier effect, the insufficient quality of the data does not allow us to proof the validity of this model. Thus, further investigations of this effect will be necessary.

To summarize, we observe a strong geometrical dependence of both the Peltier and the Seebeck effect in graphene nanoconstrictions dominating over the previously reported thermoelectric effect at the graphene-metal interface.[4,20] We can explain this local variation of the Seebeck coefficient by a reduction in the EMFP, which is caused by the increased scattering from the edges. Compared to Au nanowires, where a similar effect has been observed previously,[15] graphene is a more suitable system for achieving control of the mean free path, due to its lower dimensionality and also comparatively bigger electron mean free path. Furthermore, we observe an additional contribution to the Peltier effect by an 'electron wind' resulting from the high drift velocity of charge carriers in the constriction. This work highlights the major influence of disorder and geometry on thermoelectric properties of graphene. Thus, thermoelectric effects are likely present in graphene whenever edge scattering becomes appreciable and can lead to undesired heating/cooling. Similarly, any temperature gradient across an edge scattering region will create a parasitic voltage drop over the device. These are important consideration for future photothermoelectric as well as thermal and electrical transport measurements in nanoscale electronic devices.

In addition, our findings have implications for thermal management in future integrated circuits made out of graphene: The results open a path to producing a single material thermocouple or Peltier element that can be precisely positioned using electron beam lithography. As shown in Figure S13 in the Supporting Information, a substantial reduction of the channel width effectively creates a highly localized Peltier element which could be used for local cooling or temperature sensing. Such all-graphene thermocouples could be integrated into planar device structures on a wafer scale and at comparatively low costs.

**Methods**

*Device fabrication*

The devices were fabricated by transferring two different types of CVD graphene,[30] multilayer (2-4 layers) and single layer (see Supporting Information), on top of a Si chip with a 300 nm $SiO_2$ and pre-patterned Cr/Au contacts using a standard wet transfer method.[31] Subsequently, the graphene was patterned into the different geometries employing standard electron-beam lithography and then etched into different geometries using oxygen plasma etching.

*Scanning Thermal Microscopy measurement methods*

The SThM is located in a high vacuum environment, prohibiting parasitic heat transfer between the tip and the sample to achieve a better thermal resolution.[18,32] In our measurements, the spatial resolution is limited by the size of the tip-sample contact which is on the order of tens of nanometers.

We used two distinct scanning measurement methods, passive SThM temperature probing and active heated-probe local thermovoltage measurements.

In the Peltier measurement, the device is electrically excited with an AC bias $V_{bias}$ through the global contacts at a frequency of $f = 17Hz$. The SThM tip is scanned over the sample, measuring the temperature $\Delta T_{peltier}$ at the first harmonic ($f$) using a SRS830 lock-in (see Figure 1a). Simultaneously the unmodulated temperature-dependent DC signal and the Joule heating signal $\Delta T_{joule}$, measured at the second harmonic (2$f$), are recorded. The Peltier and Joule measurements were performed following Menges *et al.*,[18] to exclude tip-sample contact-related artefacts (see section 7 in the Supporting Information and [18]).

In contrast, for the thermovoltage scanning method, the SThM tip is heated up by applying a high AC voltage of $V_{heater}$ = 2.24 $V_{pp}$ to the temperature sensor. This Joule heating of the SThM tip at a frequency of $f_T$ = 57Hz, results in a modulation of the SThM resistor temperature of approximately 60K, leading to a SThM tip temperature modulation of $\Delta T \approx 18 \pm 2$ K at the interface with graphene (see section 7 Supporting Information). This local heat source is then scanned over the sample while the global voltage drop $V_{th}$ over the two contacts is measured with a SR560 voltage pre-amplifier and a SRS830 lock-in amplifier at the second harmonic ($2f_T$) (see Figure 1b). Our thermovoltage measurements do not require electrical contact between the tip and the sample, as does a similar method reported previously,[33] and thereby eliminate linked uncertainty, as well as requirements on the strength of the electrical tip-sample contact. To rule out effects on the measured signal stemming from accidental phase errors in the lock-in signal, we performed a DC measurement where a positive and negative square wave are applied respectively and the two resulting temperature maps are subtracted. This configuration shows the same signal as the AC measurements, thereby eliminating the possibility of an unintended phase effect causing the signal (see Supporting Information).

ASSOCIATED CONTENT

**Supporting Information**.

The Supporting Information is available free of charge from the ACS website and DOI. Peltier correction terms, graphene characterization, EMFP evaluation, *I-V* traces, DC measurement, thermovoltage measurements, SThM setup, FEM analysis, current dependencies other geometries, long ribbon geometry, spatially dependent Seebeck coefficient theory, gate dependency, heat movement by carriers, island geometry measurements, sample preparation. (PDF)

AUTHOR INFORMATION

**Corresponding Author**

*Email: p.gehring@udelft.nl; o.kolosov@lancaster.ac.uk

**Author Contributions**

∥ A. H. and J. S. contributed equally to this work.

A.H. fabricated the devices. A.H., P.G. and O.K. drafted the manuscript. A.H. J.S., C.E. and P.G. performed the measurements. A.H. J.S., C.E., P.G., J.M. and O.K. analyzed and processed the data. Y.S. and J.W. provided the graphene. E.M. carried out the theoretical modelling and J.S. performed the FEM calculations. O.K. and P.G. supervised, conceived and designed the experiments. All authors discussed the results and contributed to manuscript revision.

**Funding Sources**

This work is supported by the UK EPSRC (grant nos. EP/K001507/1, EP/J014753/1, EP/H035818/1, EP/K030108/1, EP/J015067/1, and EP/N017188/1).

**Notes**

The authors declare no competing financial interest.

ACKNOWLEDGMENT

The authors would like to thank Prof. Colin Lambert and Dr. Hatef Sadeghi for helpful discussions and Jasper Fried for performing the long ribbon conductance measurements. P.G. acknowledges a Marie Skłodowska-Curie Individual Fellowship under grant TherSpinMol (ID: 748642) from the European Union's Horizon 2020 research and innovation programme. O.K. acknowledges the EPSRC project EP/K023373/1 and EU project No.604668 QUANTIHEAT.

ABBREVIATIONS

SThM, scanning thermal microscope; EMFP, electron mean free path; AFM, atomic force microscopy.

(26) Li, Q.; Das Sarma, S. Finite temperature inelastic mean free path and quasiparticle lifetime in graphene. *Phys. Rev. B* **2013**, *87,* 085406.

(27) Castro Neto, A. H.; Guinea, F.; Peres, N. M. R.; Novoselov, K. S.; Geim, A. K. The electronic properties of graphene. *Rev. Mod. Phys.* **2009**, *81*, 109-162.

(28) Zuev, Y. M.; Chang, W.; Kim, P. Thermoelectric and Magnetothermoelectric Transport Measurements of Graphene. *Phys. Rev. Lett.* **2009**, *102,* 096807.

(29) Bae, M.-H.; Islam, S.; Dorgan, V. E.; Pop, E. Scaling of High-Field Transport and Localized Heating in Graphene Transistors. *ACS Nano* **2011**, *5*, 7936-7944.

(30) Sheng, Y.; Rong, Y.; He, Z.; Fan, Y.; Warner, J. H. Uniformity of large-area bilayer graphene grown by chemical vapor deposition. *Nanotechnology* **2015**, *26*, 395601.

(31) Li, X.; Cai, W.; An, J.; Kim, S.; Nah, J.; Yang, D.; Piner, R.; Velamakanni, A.; Jung, I.; Tutuc, E.; Banerjee, S. K.; Colombo, L.; Ruoff, R. S. Large-Area Synthesis of High-Quality and Uniform Graphene Films on Copper Foils. *Science* **2009**, *324*, 1312-1314.

(32) Kim, K.; Jeong, W.; Lee, W.; Reddy, P. Ultra-High Vacuum Scanning Thermal Microscopy for Nanometer Resolution Quantitative Thermometry. *ACS Nano* **2012**, *6*, 4248-4257.

(33) Lee, B.; Kim, K.; Lee, S.; Kim, J. H.; Lim, D. S.; Kwon, O.; Lee, J. S. Quantitative Thermopower Profiling across a Silicon p–n Junction with Nanometer Resolution. *Nano Lett.* **2012**, *12*, 4472-4476.
21

**Graphical TOC Entry**

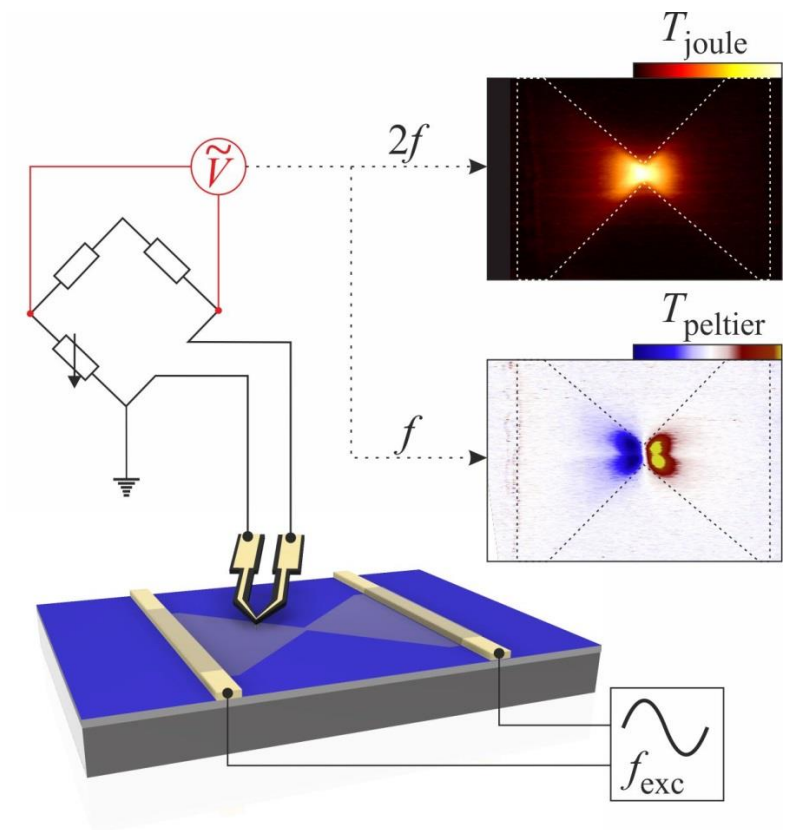